\documentclass[a4paper]{ifacconf}

\usepackage{amsmath}
\usepackage{amsfonts}
\usepackage{amssymb}
\usepackage{exscale}
\usepackage{graphicx}
\usepackage{color}

\usepackage{algorithmic}
\usepackage{graphicx}
\usepackage{gensymb}
\usepackage{multirow}
\usepackage{array}
\usepackage{makecell}
\usepackage[utf8]{inputenc}
\usepackage{url}
\usepackage{epsfig}
\usepackage{lscape}
\usepackage{todonotes}								
\usepackage{subeqnarray}
\usepackage{calc,array}
\usepackage{upgreek}
\usepackage{bm}
\usepackage{natbib}        
\usepackage{fancybox}
\usepackage{booktabs}

\usepackage{epstopdf}
\newtheorem{theorem}{Theorem}
\usepackage{array}

\usepackage{textcomp} 
\usepackage{mathdots}
\usepackage{siunitx}
\usepackage{import}

\makeatletter
\def\endfigure{\end@float}
\def\endtable{\end@float}
\makeatother

\let\ifacconfcaptionwidth\captionwidth
\usepackage[caption=false]{subfig}
\let\captionwidth\ifacconfcaptionwidth

\usepackage{tikz}
\usetikzlibrary{arrows, positioning, calc}
\usepackage{environ}

\makeatletter
\newsavebox{\measure@tikzpicture}
\NewEnviron{scaletikzpicturetowidth}[1]{%
	\def\tikz@width{#1}%
	\begin{lrbox}{\measure@tikzpicture}%
		\BODY
	\end{lrbox}%
	\pgfmathparse{#1/\wd\measure@tikzpicture}%
	\BODY
}
\makeatother

\renewcommand{\j}{\mathrm{j}}

\renewcommand{\Re}{\text{Re}}

\begin{document}
\begin{frontmatter}
\title{Fractional-Order Partial Cancellation of Integer-Order Poles and Zeros}

\author[First]{Benjamin Voß}
\author[First]{Christoph Weise}
\author[Third]{Michael Ruderman}
\author[First]{Johann Reger\thanksref{Fourth}} 

\address[First]{Control Engineering Group, Technische
 Universit\"at Ilmenau, P.O. Box 10 05 65, D-98684, Ilmenau, Germany}%
\address[Third] {Department of Engineering Sciences, University of Agder,
  P.O. Box 422, 4604 Kristiansand, Norway, {\tt\small michael.ruderman@uia.no}}

\thanks[Fourth]{Corresponding author: {\tt\small johann.reger@tu-ilmenau.de}\\ \copyright ~2022 the authors. This work has been accepted to IFAC for publication under a Creative Commons Licence CC-BY-NC-ND.}%

\begin{abstract}
The key idea of this contribution is the partial compensation of non-minimum phase zeros or unstable poles. Therefore the integer-order zero/pole is split into a product of fractional-order pseudo zeros/poles. The amplitude and phase response of these fractional-order terms is derived to include these compensators into the loop-shaping design. Such compensators can be generalized to conjugate complex zeros/poles, and also implicit fractional-order terms can be applied. In the case of the non-minimum phase zero, its compensation leads to a higher phase margin and a steeper open-loop amplitude response around the crossover frequency resulting in a reduced undershooting in the step-response, as illustrated in the numerical example.
\end{abstract}

\end{frontmatter}
\section{Introduction}
Non-minimum phase zeros and unstable poles limit the closed-loop performance of control systems. Regarding the open-loop controller design, these dynamics restrict the achievable crossover frequency and stability margins. For these processes, the idea of direct compensation does not render the control loop internally stable and bounded disturbances destabilize the system. Introducing fractional-order (FO) terms in the controller, i.e. $s^\alpha$ in the Laplace domain, however, enables one to split the undesired term 
and set up an easy to tune partial compensator.
 
In this paper we reflect on the results published in \cite{MerrikhBayat2013} and extend the observations therein by extending the proposed method towards conjugate complex zeros/poles. Furthermore, we include the implicit FO zero/pole for partial cancellation of the undesired integer-order (IO) non-minimum phase zeros or unstable poles, i.e. the individual parts of the FO lead-lag-element, usually utilized to achieve the iso-damping property \citep{Tavazoei2014,Raynaud2000}. The FO compensators are straightforward to tune and can be implemented using IO approximations. Comparable results can be achieved with advanced IO design methods leading to a similar controller order, see e.g. \citep{Voss2022a}.

This contribution is divided into four sections. The following section gives a short introduction to FO transfer functions and introduces the analytical amplitude- and phase responses of FO pseudo zeros/poles and its implicit counterparts.
In Section~\ref{sec:application} we split an IO zero/pole into the corresponding FO pseudo zeros/poles and partially compensate the stable part, leading to changes in the corresponding sensitivities. Finally, in Section~\ref{sec:examples}, the proposed method 
is evaluated using an academic example,
showing its benefit in comparison to stable-mirror compensation with an IO zero/pole. 
Section~\ref{sec:conclusions} concludes the paper.

\section{Preliminary Results and Definitions}
\label{sec:fundamentals}
\subsection{Fractional-Order Control}
The generalization of integer-order calculus towards non-integer orders can be done in various ways. 
A common operator applied in control theory is the non-local Caputo operator \citep{MonjeCVXF10}
\begin{equation}
\label{eq:Caputo-definition}
{}_{t_0}\mathcal{D}^{\alpha}_t f(t) = \frac{1}{\Gamma(m-\alpha)} \int_{t_0}^{t} \dfrac{f^{(m)}(t)}{(t-\tau)^{\alpha-m+1}} \mathrm{d}\tau
\end{equation}
where $\alpha\in \mathbb{R}^+$ is the order of differentiation, $m$ is an
integer such that $m-1 \leq \alpha<m$ and $\Gamma(\cdot)$ represents Euler's gamma function.
The operator's Laplace transform is given by \citep{MonjeCVXF10}
\vspace{-1.2ex}
\begin{equation}
\label{eq:Caputo-laplacetransform}
\mathcal{L}\left\{{}_0\mathcal{D}_t^{\alpha} f(t) \right\} = s^{\alpha}\mathcal{L}\{f(t)\} - \sum\limits_{k=0}^{m-1}s^{\alpha-k-1}f^{(k)}(0).
\end{equation}
Here we shall only consider FO input-output behavior with zero initial conditions. The transfer function reads
\begin{equation}
\label{eq:generaltffunction}
G(s) = 
\dfrac{b_ms^{m \alpha}+\hdots+b_1s^{\alpha}+b_0}{a_ns^{n\alpha}+\hdots+a_1s^{\alpha}+1}
= \dfrac{B(s^\alpha)}{A(s^\alpha)}
\end{equation}
with pseudo polynomials $A$, $B$ of commensurate order $\alpha$.

\begin{theorem}[Stability:\! FO\! LTI\! systems \!\citep{Matignon1996}]
	The FO system with transfer function ${G}(s) \!=\!
        B(s^\alpha)/A(s^\alpha)$ and commensurate order $\alpha$ is stable iff
        any $p\in\mathbb{C}$ of $A$ with $A(p)=0$ satisfies
	\vspace{-1.2ex}
	\begin{displaymath}
	|\arg(p)| > \alpha \dfrac{\pi}{2}.
	\end{displaymath}
\end{theorem}
Compared to IO systems, this results in a larger, however non-convex stability domain for $\alpha \in (0,1)$, see Fig.~\ref{fig:fo-stability}.

\subsection{Fractional-Order Pseudo Zeros and Poles}
\label{ssec:fundamentals-fo-exp}
We consider the term given in the Laplace domain 
\begin{equation}\label{eq:fo-zero-exp}
	X_{z,\alpha}^k(s) = \left(1- \left(\frac{s}{z}\right)^{\alpha}\right)^k,\quad z > 0,\,\alpha\in (0,1]
\end{equation}
with $k\in\{-1,1\}$, i.e. a non-minimum phase pseudo zero of non-integer order $\alpha$ for $k=1$ or an unstable pseudo pole for $k=-1$. The magnitude and phase of $X_{z,\alpha}^k$ can be calculated as (\cite{MerrikhBayat2013}):
\begin{align}
	\left|X_{z,\alpha}^k(\j\omega)\right|^2 &=\left(1 + \left(\frac{\omega}{z}\right)^{2\alpha} - 2\left(\frac{\omega}{z}\right)^\alpha \cos \left(\frac{\pi\alpha}{2}\right)\right)^k ,\\
	\angle X_{z,\alpha}^k(\j\omega) &= k\arctan\left( \frac{-\left(\frac{\omega}{z}\right)^\alpha\sin \left(\frac{\pi\alpha}{2}\right)}{1- \left(\frac{\omega}{z}\right)^\alpha\cos\left(\frac{\pi\alpha}{2}\right)}\right) .\label{eq:fo-zero-expl-phase}
\end{align}
Note that \cite{MerrikhBayat2013} limit this term to ${z\in\mathbb{R}}$, but a straightforward generalization to $z\in\mathbb{C}$ is possible. 

For clarity, we restrict the discussion in this section to $k=1$, i.e. right-half plane (RHP) pseudo zeros. To derive analog results for RHP pseudo poles, just let $k=-1$.

\begin{figure}[ht]
	\centering
	\includegraphics[width=1\linewidth]{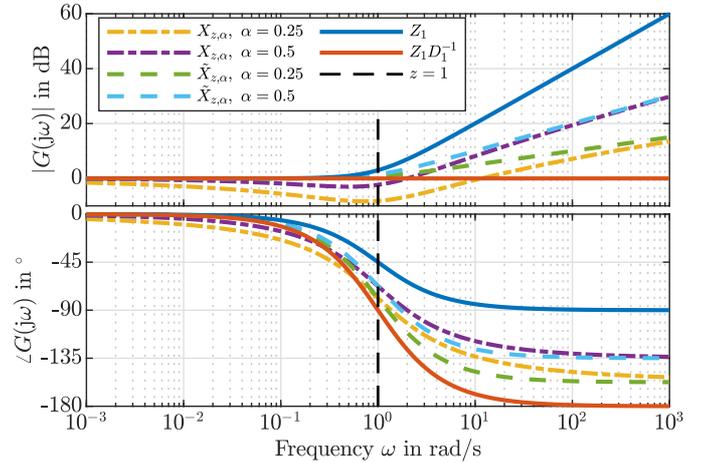}
	\caption{Bode plot of $X_{z,\alpha}$ and $\tilde{X}_{z,\alpha}$ opposite to the IO term~${Z_1}$ and its pseudo compensation $Z_1D_1^{-1}$.}
	\label{fig:fo-zero-bode}
\end{figure}

The Bode plot of $X_{z,\alpha}$ for $z = 1$ and ${\alpha\in \{0.25,\,0.5,\,1\}}$ is depicted in Fig.~\ref{fig:fo-zero-bode}. Note that $X_{z,\alpha}$ of~\eqref{eq:fo-zero-exp} coincides with an IO non-minimum phase zero for $\alpha=1$, i.e.
\begin{equation}
	Z_1(s) = 1-\frac{s}{z},\quad z>0
\end{equation} 
with $Z_1 = X_{z,1}$, which we introduce to clearly identify IO zeros.
The asymptotic behavior for low and high frequencies is summarized in Table~\ref{tab:do-zero-asymptotics}.
It shows that for high frequencies the magnitude slope decreases for small values of $\alpha$. This does not apply to the phase lag, which increases for a decreasing order $\alpha$. 
Compared to the IO case, the phase drop already occurs at lower frequencies. Furthermore, the amplitude response of~$X_{z,\alpha}$ shows a minimum at the frequency 
\citep{MerrikhBayat2013}
\begin{align*}
	\omega_\mathrm{min} = z\left(\cos\left(\frac{\pi\alpha}{2}\right)\right)^{\frac{1}{\alpha}} \quad\text{with}		\\
	\left|X_{z,\alpha}^1(\j\omega_{\mathrm{min}})\right| = \sin\left(\frac{\pi\alpha}{2}\right),\quad
	 \angle  X_{z,\alpha}^1(\j\omega_{\mathrm{min}}) &= \frac{\pi}{2}\left( \alpha-1\right).
\end{align*} 

\begin{figure}[ht]
	\centering
	\vspace{-.8ex}
	\includegraphics[width=.6\linewidth]{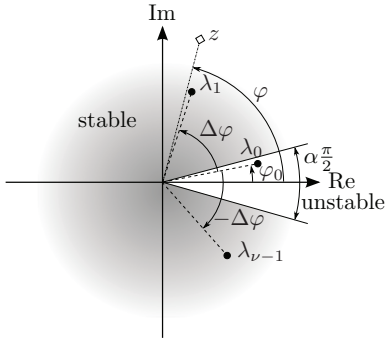}
	\caption{Schematic representation of the stability condition.}
	\label{fig:fo-stability}
\end{figure}

Given an IO transfer function of a plant containing a non-minimum phase zero or unstable pole at $z>0$
\begin{equation}\label{eq:fo-zero-exp-plant}
	G_1(s) = \left(1-\frac{s}{z}\right)^k \hat{G}_1(s) =Z_1^k(s) \hat{G}_1(s),
\end{equation}
$Z_1$ can be expanded to $\alpha^{-1}=\nu,\, \nu\in\mathbb{N}$ pseudo zeros (\cite{MerrikhBayat2013}), leading to
\begin{equation}\label{eq:fo-exp-expansion}
	Z_1^k(s) = \underbrace{\left(1-\left(\frac{s}{z}\right)^{\frac{1}{\nu}}\right)^k}_{X_{z,\nu^{-1}}^k(s)}~
	\underbrace{\left(\sum_{n=1}^\nu \left( \frac{s}{z}\right)^{\frac{n-1}{\nu}}\right)^k}_{Q_{z,\nu}^k(s)}.
\end{equation}
Note that all $\nu-1$ pseudo zeros of $Q_{z,\nu} $ are located in the stable region $\mathbb{C}_\alpha=\{z\in\mathbb{C}\,\left|\,\left|\arg(z)\right|\right.>\alpha\frac{\pi}{2}\}$, depicted in Fig.~\ref{fig:fo-stability}, see \citep{MerrikhBayat2013}. 
Thus, it may be used to partially cancel the non-minimum phase zero of the plant
\begin{equation}\label{eq:fo-cancellation-exp}
	Z_1^k(s)~Q_{z,\nu}^{-k}(s) =\left( 1-\left(\frac{s}{z}\right)^{\frac{1}{\nu}}\right)^k=  X_{z,\nu^{-1}}^k(s),
\end{equation}
leading to a FO pseudo zero described above for $k=1$. Note, although the phase lag of $X_{z,\nu^{-1}}$ exceeds $Z_1$ (see Fig.~\ref{fig:fo-zero-bode}), the partial cancellation results in a greater phase margin as well as a steeper slope of the magnitude in comparison with the classical IO pseudo compensation of the non-minimum phase zero by its mirrored pole
\begin{equation}\label{eq:fo-def-D1}
D_1^k(s) = \left(1+\frac{s}{z}\right)^{k}.
\end{equation}

\begin{table}[ht]
	\centering
	\caption{Asymptotic frequency characteristics of (implicit) FO pseudo zeros and poles.}
	\label{tab:do-zero-asymptotics}
	\begin{tabular}{r|ll|ll}
														& \multicolumn{2}{l|}{\textbf{Pseudo Zero}}										& \multicolumn{2}{l}{\textbf{Pseudo Pole}} \\\hline
		Frequency									&  	$\omega \ll z$			&$\omega \gg z$			&  	$\omega \ll z$			&$\omega \gg z$  \\
	 	 $\frac{\mathrm d}{\mathrm d\omega}|\cdot|$ in $\frac{\mathrm{dB}}{\mathrm{dec}}$	&$0$	&$20\,\alpha$&$0$	&$-20\,\alpha$ \\
	  $\angle(\cdot) $ in $^\circ$ & $ 0$ 		 &$  -180 + 90\,\alpha$ & $ 0$ 		 &$  180 - 90\,\alpha$\\
	\end{tabular}
\end{table}

\subsection{Implicit Fractional-Order Pseudo Zeros and Poles}
\label{ssec:fundamentals-fo-imp}
An effect similar to the partial cancellation of the non-minimum phase zero in~\eqref{eq:fo-cancellation-exp} can be achieved by introducing the implicit term as a part of the FO lead-lag element \citep{MonjeCVXF10}
\begin{equation}\label{eq:fo-Q-imp}
	\tilde Q_{z,\nu}^k(s) = \left(1+\frac{s}{z}\right)^{k\frac{\nu-1}{\nu}},\quad \nu \in \mathbb{N}
\end{equation}
with the IO pseudo compensation as an edge case, i.e.
$\displaystyle\lim_{\nu\rightarrow\infty}\tilde{Q}_{z,\nu}^{k}(s) = D_1^k(s)$.
The term~$\tilde Q_{z,\nu}^k$ leads to the implicit counterpart of~\eqref{eq:fo-zero-exp}:
\begin{equation}\label{eq:fo-zero-imp}
	\tilde X_{z,\alpha}^k(s) = \left(1- \frac{s}{z}\right)^k~\tilde Q_{z,\nu}^{-k}(s),\quad z > 0,\,\alpha = \frac{1}{\nu}
\end{equation}
and $k\in\{-1,1\}$. The magnitude and phase
\begin{align}
	\left|\tilde X_{z,\alpha}^k(\j\omega)\right| 		\! &= \! \left(1+\left(\frac{\omega}{z}\right)^2\right)^{k\frac{\alpha}{2}},\\
	\angle \tilde X_{z,\alpha}^k(\j\omega) \!	&=\! k\arctan\left(-\frac{\omega}{z}\right) - k(1-\alpha)\arctan\left(\frac{\omega}{z}\right) \label{eq:fo-zero-impl-phase}
\end{align}
can be calculated directly from the IO terms. 

Again, we restrict the following discussions to implicit pseudo zeros, i.e. $k=1$. However, the analog results for implicit pseudo poles are obtained by setting $k=-1$.

The asymptotic frequency characteristics coincide with the pseudo zero in \eqref{eq:fo-zero-exp} and are given in Table~\ref{tab:do-zero-asymptotics}. Furthermore, the phase of the two representations in~\eqref{eq:fo-zero-expl-phase} and~\eqref{eq:fo-zero-impl-phase} coincide at $ \omega = z $, i.e.
\begin{equation}
	\angle  X_{z,\alpha}^k(\j\omega)\Big|_{z} = \angle \tilde X_{z,\alpha}^k(\j\omega)\Big|_{z} = k~\frac{\pi}{2}\left(\frac{\alpha}{2}-1 \right).
\end{equation} 
However, the sample graphs 
in Fig.~\ref{fig:fo-zero-bode} show the major differences between these two formulations. First, the magnitude of the implicit term $\tilde X_{z,\alpha}$ does not lower around $\omega = z$. Second, it results in less phase lag for $\omega<z$ compared to the explicit formulation in~\eqref{eq:fo-zero-exp}. This renders the implicit term $\tilde Q_{z,\nu}$ more attractive to partially compensate a non-minimum phase zero of a given plant, especially in the case of phase limitations.

\subsection{Pair of FO Conjugate Complex Pseudo Zeros or Poles}\label{ssec:fundamentals-fo-z2}
The practical relevance of RHP conjugate complex zeros might not be as obvious as it is for poles. However, consider the Padé-approximation of a time delay. It involves a dominant pair of non-minimum phase zeros for approximation orders higher than one \citep{Pade1892}.
On the other hand, given a plant with a pair of conjugate complex poles, undesired effects may occur not only for those in the RHP. Also low-damped stable poles have a significant impact on the stability margins. So first we have a look at RHP zeros and poles, then a pair of stable poles is considered.

Similar to the representations \eqref{eq:fo-zero-exp} to~\eqref{eq:fo-zero-expl-phase}, a pair of conjugate complex pseudo zeros at 
\begin{equation}\label{eq:fo-def-z12}
	z =\omega_0 e^{\j\varphi},\quad \bar{z} =\omega_0 e^{-\j\varphi}\in\mathbb{C}_+ 
\end{equation} 
with $\omega_0=|z|$, $\varphi=\arg(z)$ and $\mathbb{C}_+=\{z\in\mathbb{C}\,\left|\,\Re(z)\right. >0\}$ may be given as $\mathcal{X}_{z,\alpha}^k  =  X_{z,\alpha}^k~X_{\bar{z},\alpha}^k $ and we obtain
\begin{equation}
\mathcal{X}_{z,\alpha}^k (s)	= \omega_0^{-2\alpha k}\left( s^{2\alpha} - 2\left(s\omega_0\right)^{\alpha} \cos\left( \varphi \alpha\right) + \omega_0^{2\alpha}\right)^k.\label{eq:fo-zero2-exp}
\end{equation} 
For $\alpha=1$, Equation~\eqref{eq:fo-zero2-exp} coincides with the IO case, i.e.
\begin{equation}\label{eq:io-zero2}
Z_2^{k}(s)  =  \omega_0^{-2k}\left( s^{2} - 2s\omega_0 \cos\left( \varphi\right) + \omega_0^{2}\right)^{k},
\end{equation} 
which can be pseudo compensated analogous to~\eqref{eq:fo-def-D1} with
\begin{equation}\label{eq:io-def-D2}
D_2^{k}(s)  =  \omega_0^{-2k}\left( s^{2} + 2s\omega_0 \cos\left( \varphi\right) + \omega_0^{2}\right)^{k}.
\end{equation} 
\begin{figure}[ht]
	\centering
	\includegraphics[width=1.0\linewidth]{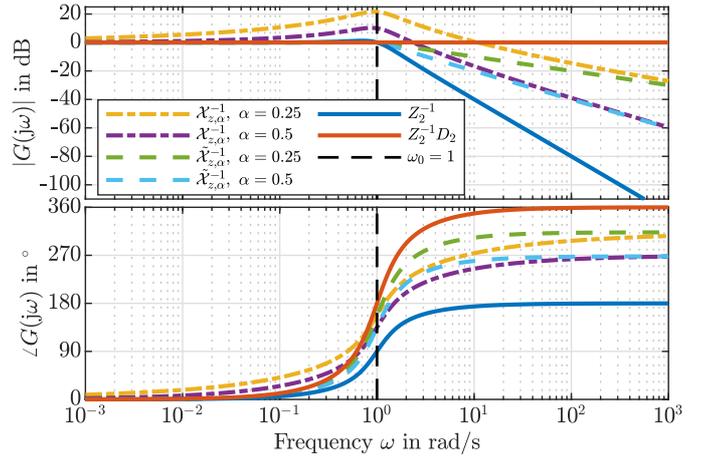}
	\caption{Bode plot of $\mathcal{X}_{z,\alpha}^{-1}$ and $\tilde{\mathcal{X}}_{z,\alpha}^{-1}$  opposite to the IO term~$Z_2^{-1}$ and its pseudo compensation $Z_2^{-1}D_2$.}
	\label{fig:fo-zero2-bode}
\end{figure}

The Bode plot of $\mathcal{X}_{z,\alpha}^{-1}$ for $z =  1e^{\j\frac{\pi}{3}}$ and ${\alpha\in \{0.25,\,0.5,\,1\}}$ is given in Fig.~\ref{fig:fo-zero2-bode}. In contrast to the case of a single positive pseudo zero, an increase in the magnitude leading to a maximum can be observed. Moreover, the phase lifts at lower frequencies compared to the IO case.  The asymptotic frequency characteristics are summarized in Table~\ref{tab:fo-zero2-asymptotics}.

\begin{table}[ht]
	\centering
	\caption{Asymptotic freq. char. of FO RHP conjugate complex pseudo zeros and poles.}
	\label{tab:fo-zero2-asymptotics}
	\setlength\tabcolsep{4pt}
	\begin{tabular}{r|ll|ll}
																	& \multicolumn{2}{l|}{\textbf{Pair of Pseudo Zeros}}										& \multicolumn{2}{l}{\textbf{Pair of Pseudo Poles}} \\\hline
		Frequency									&  	$\omega \ll \omega_0$			&$\omega \gg \omega_0$			&  	$\omega \ll \omega_0$			&$\omega \gg \omega_0$  \\
	 $\frac{\mathrm d}{\mathrm d\omega}|\cdot|$ in $\frac{\mathrm{dB}}{\mathrm{dec}}$	&$0$	&$40\,\alpha$	&$0$	&$-40\,\alpha$ \\
	  $\angle(\cdot) $ in $^\circ$& $ 0$ 		 &$  -360 + 180\alpha$ & $ 0$ 		 &$  360 - 180\alpha$\\
	\end{tabular}
\end{table}

Given a plant with a pair of RHP conjugate complex zeros ($k=1$) or poles ($k=-1$), it can be written as
\begin{equation}
	G_2(s) = Z_2^k(s)\hat{G}_2(s).
\end{equation}
Equation~\eqref{eq:io-zero2} can be reformulated as
\begin{equation}\label{eq:io-zero-complex}
	Z_2^k(s) =  \left(1-\frac{s}{z}\right)^k  \left(1-\frac{s}{\bar{z}}\right)^k
\end{equation} 
which turns out to be useful since expansion~\eqref{eq:fo-exp-expansion} also holds for $z\in \mathbb{C}$.
Thus, each term in~\eqref{eq:io-zero-complex} can be expanded to~\eqref{eq:fo-exp-expansion},
where $X_{z,\nu^{-1}}^k$ and $Q_{z,\nu}^k$ have a conjugate complex part, i.e. 

\begin{equation}\label{eq:fo-Q2-exp}
	\mathcal{Q}_{z,\nu}^k(s)= Q_{z,\nu}^k(s)~Q_{\bar{z},\nu}^k(s)
\end{equation}
with real coefficients. In order to make use of $ \mathcal{Q}_{z,\nu}^k$, we show that all pseudo zeros are located in the stable region~$\mathbb{C}_\alpha$ and restrict us to $k=1$ (pseudo zeros) without loss of generality. 
For this purpose, we consider one term of $\mathcal{Q}_{z,\nu}$ in~\eqref{eq:fo-Q2-exp} separately and its associated IO zero in~\eqref{eq:io-zero-complex}. Then the same discussion is made for the other term.

The arguments of the $\nu$-th roots $\lambda_n = \sqrt[\nu]{z}$ are given by
\begin{equation}\label{eq:fo-arg-nuth-root}
\arg\left(\lambda_n\right) = \varphi_n = (\varphi + 2\pi n) \nu^{-1},\quad n = 0,1,\ldots,\nu-1
\end{equation}
with the constant angle between two consecutive roots 
\begin{equation}\label{eq:fo-deltaphi}
\Delta \varphi = \varphi_n - \varphi_{n-1} = 2\pi \nu^{-1}.
\end{equation}
From expansion~\eqref{eq:fo-exp-expansion} it can be seen that the principal root~$\lambda_0$ is separated and not part of $Q_{z,\nu}$, leading to the zeros of $Q_{z,\nu}$ with
\begin{equation}
\varphi_n = (\varphi + 2\pi n)\nu^{-1} ,\quad \quad n = 1,2\ldots,\nu-1.
\end{equation}
Considering the two critical cases that are illustrated in Fig.~\ref{fig:fo-stability}, i.e. $\lambda_1$ and $\lambda_{\nu-1}$, we need to show that
\begin{equation}
\Delta \varphi > \pi\nu^{-1},
\end{equation}
as it covers the worst cases for  $|\varphi|<\frac{\pi}{2\nu}$, i.e. $\varphi = \pm \frac{\pi}{2}$. It directly follows from~\eqref{eq:fo-deltaphi} that this holds true
and thus all pseudo zeros of $\mathcal{Q}_{z,\nu}$ are located in the stable region.

Therefore, without introducing instability, it can be used to partially cancel the pair of conjugate complex zeros/poles resulting in
\begin{equation}
	Z_2^k(s)~\mathcal{Q}_{z,\nu}^{-k}(s) = \mathcal{X}_{z,\alpha}^k(s)
\end{equation}
with a lower-order pair of pseudo zeros/poles.

Analogous to \eqref{eq:fo-Q-imp} and~\eqref{eq:fo-zero-imp}, the implicit counterparts of $\mathcal{Q}_{z,\nu}^k$ and $\mathcal{X}_{z,\alpha}^k$ can be defined as
\begin{align}
	\tilde{\mathcal{Q}}_{z,\nu}^k(s) &= \left(\frac{1}{\omega_0^{2}}\left( s^{2} + 2s\omega_0 \cos\left( \varphi\right) + \omega_0^{2}\right)\right)^{k\frac{\nu-1}{\nu}}, \label{eq:fo-zero2-impl-Q}\\
	\tilde{\mathcal{X}}_{z,\alpha}^k(s) &= Z_2^k(s)~\tilde{\mathcal{Q}}_{z,\nu}^{-k}(s) \nonumber\\
		&= \left( \omega_0^{2\alpha}\frac{s^{2} - 2s\omega_0\cos(\varphi) + \omega_0^{2}}{\big(s^{2} + 2s\omega_0\cos(\varphi) + \omega_0^{2}\big)^{1-\alpha}} \right)^k\label{eq:fo-zero2-impl}
\end{align}
with $z,\bar{z}$ of~\eqref{eq:fo-def-z12} and $\alpha = \frac{1}{\nu}$. Similar to the single implicit pseudo zero/pole $\tilde X_{z,\alpha}^k$ in~\eqref{eq:fo-zero-imp}, the magnitude and phase of $\tilde{\mathcal{X}}_{z,\alpha}^k$ can be calculated using the IO terms (omitted here). Figure~\ref{fig:fo-zero2-bode} holds an exemplary Bode plot of $\tilde{\mathcal{X}}_{z,\alpha}^{-1}$. Asymptotic frequency characteristics are summarized in Table~\ref{tab:fo-zero2-asymptotics}. Comparing the behavior of the explicit and implicit pair of pseudo poles, similar conclusions can be drawn as for a single real non-minimum phase zero.

Interesting observations can be made when considering a pair of complex poles in ${\mathbb{C}_-=\{z\in\mathbb{C}\,\left|\,\Re(z)\right. <0\}}$, i.e.
\begin{equation}\label{eq:fo-def-p12}
	p = \omega_0 e^{\j \varphi},\quad \bar{p} = \omega_0 e^{-\j \varphi}\in\mathbb{C}_-,
\end{equation}
as the partial cancellation is not necessarily restricted to RHP poles. For this purpose, we show that all $\nu=\alpha^{-1}$, $\nu\in\mathbb{N}$ roots of $p$ and $\bar{p}$ are located in the stable region $\mathbb{C}_{\alpha}$, that is
\begin{equation}\label{eq:fo-pole2-roots-stability}
	\left| \arg(p)\right| > \frac{\pi}{2} \quad \Longrightarrow\quad p^{\alpha} \in \mathbb{C}_\alpha,~ \alpha = \nu^{-1}.
\end{equation} 
Equation~\eqref{eq:fo-arg-nuth-root} for $\lambda_n = \sqrt[\nu]{p}$ is equivalent to
\begin{equation}
	\nu\arg\left(\lambda_n\right) - 2\pi n = \varphi,\quad n = 0,1,\ldots,\nu-1.
\end{equation}
Considering the absolute values with the restricted range ${0<\left|\arg(\lambda_n)\right|<\pi}$ results in
\begin{equation}
\nu \left| \arg\left(\lambda_n\right) \right| \geq | \varphi | \quad \Longleftrightarrow \quad  \left| \arg\left(\lambda_n\right) \right| \geq \frac{| \varphi |}{\nu}
\end{equation}
and for $| \varphi | >\pi/2$ we get
\begin{equation}
\left|\arg(p)\right| = | \varphi |> \frac{\pi}{2}\quad \Longrightarrow  \quad\left| \arg\left(\lambda_n\right) \right| > \frac{\pi}{2\nu} = \alpha\frac{\pi}{2}. 
\end{equation}
Therefore, implication~\eqref{eq:fo-pole2-roots-stability} holds.

\begin{figure}[ht]
	\centering
	\includegraphics[width=1.0\linewidth]{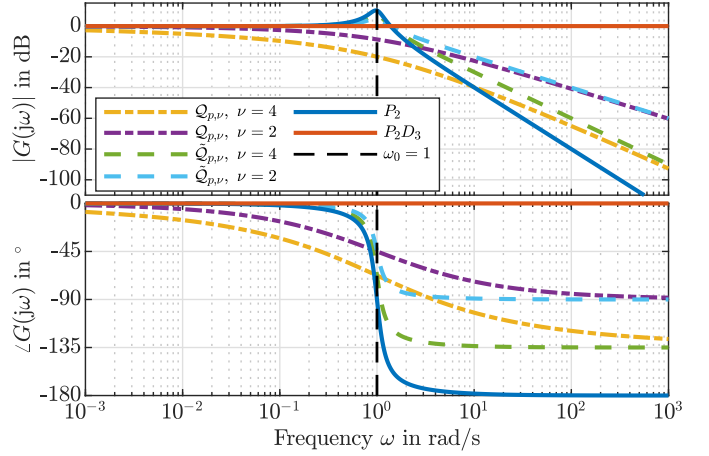}
	\caption{Bode plot of $\mathcal{Q}_{p,\nu}$ and $\tilde{\mathcal{Q}}_{p,\nu}$  opposite to the IO term~${P}$ and its compensation $P D_3$ with ${D_3 = P^{-1}}$.}
	\label{fig:fo-pole2-stable-bode}
\end{figure}

As all $\nu$-th roots are stable, any pair of conjugate complex pseudo poles could be canceled partially. However, the highest effect is expected for the roots closest to the stability border because pseudo poles with $\left|\arg(\lambda)\right|<\frac{\pi}{\nu}$ lead to an oscillating step response, see (\cite{MonjeCVXF10}). This applies to the principal roots
\begin{equation}
	\lambda_0  = \sqrt[\nu]{\omega_0}~e^{\j\frac{\varphi}{\nu}}\quad\text{and}\quad
	\bar{\lambda}_0  = \sqrt[\nu]{\omega_0}~e^{-\j\frac{\varphi}{\nu}}
\end{equation} 
that can be canceled using the term
\begin{align}
	\mathcal{X}_{p,\alpha}(s) &= \left(1-\left(\frac{s}{p}\right)^{\alpha}\right)\left(1-\left(\frac{s}{\bar{p}}\right)^\alpha\right) \nonumber\\
	&=  \omega_0^{-2\alpha}\left( s^{2\alpha} - 2\left(s\omega_0\right)^{\alpha} \cos\left( \varphi \alpha\right) + \omega_0^{2\alpha}\right)
\end{align}
with $\alpha=\nu^{-1}$, which coincides with $\mathcal{X}_{z,\alpha}$ of~\eqref{eq:fo-zero2-exp} for $z=p$. Thus, given a plant with a stable conjugate complex pole pair $p$ and $\bar{p}$, i.e. $\arg(p)=\varphi > \frac{\pi}{2}$, we get
\begin{align}
	&G_3(s) = \frac{\omega_0^{2}}{ s^{2} - 2s\omega_0 \cos\left( \varphi \right) + \omega_0^{2}}\hat{G}_3(s)
	= P(s)\hat{G}_3(s)\\
&\text{and}\quad	P(s)\mathcal{X}_{p,\alpha}(s) = \mathcal{Q}_{p,\nu}(s),\label{eq:fo-pole2-cancellation-exp}
\end{align}
where $\mathcal{Q}_{p,\nu}$ coincides with $\mathcal{Q}_{z,\nu}$ in~\eqref{eq:fo-Q2-exp} for $z=p$.
A Bode plot of $\mathcal{Q}_{p,\nu}$ for $p = -1e^{\j \frac{9\pi}{20}}$ and $\nu \in \{2,4\}$  is depicted in Fig.~\ref{fig:fo-pole2-stable-bode} and clearly shows the effect of the cancellation. According to (\cite{MonjeCVXF10}), a non-oscillating step response is expected. 

The implicit representation corresponding to~\eqref{eq:fo-pole2-cancellation-exp} for ${\alpha=\nu^{-1}}$ can be written as
\begin{align}
\tilde{\mathcal{X}}_{p,\alpha}(s) &= \omega_0^{-2\alpha}\!\left( s^{2} - 2s\omega_0 \cos\left( \varphi \right) + \omega_0^{2}\right)^{\alpha} \!\!= P^{-\alpha}(s), \label{eq:fo-pole2-impl-Qprime}\\
\tilde{\mathcal{Q}}_{p,\nu}(s) &= P(s)\tilde{\mathcal{X}}_{p,\alpha}(s) = P^{\frac{1-\nu}{\nu}}(s) ,
\end{align}
leading to a similar asymptotic frequency behavior, as it can be seen in the Bode plot of Fig.~\ref{fig:fo-pole2-stable-bode}. However, there are major differences at $\omega\approx \omega_0$ for both the magnitude as well as the phase plot. As expected, explicit cancellation of the two principal roots completely erases the oscillatory part of the poles, whereas the implicit compensation only reduces the amplitude peak and phase drop.

\section{Application to Standard Control-Loop}\label{sec:application}
For analyzing effects of the partial pole/zero cancellation on the closed-loop stability, a standard control-loop 
with plant $G$ and controller $C$ is investigated. 

The complementary, output and input sensitivity functions are given by
\begin{equation}
T = \frac{GC}{1+GC},\quad
S_\mathrm{y} = \frac{1}{1+GC},\quad
S_\mathrm{u} = \frac{G}{1+GC},
\end{equation}
where the argument `$s$' is omitted for clarity. 

Let $G = Z_i^k\hat{G}$ and $C = Q_i^{-k}\hat{C}$ with $i\in\{1,2\}$, $k\in\{-1,1\}$, where
$\displaystyle X_i^k = \textstyle\prod_{m=1}^i X_{z,\alpha}^k$ and $\displaystyle  Q_i^k = \textstyle\prod_{m=1}^i Q_{z,\nu}^k$. If $i=1$, we consider a real $z>0$, whereas $z,\bar{z}\in\mathbb{C}_+$ for $i=2$, which is consistent with the definitions in the previous sections. 
Furthermore, we stress that all RHP-roots of $GC$ are part of $Z_i^k$. Now, consider the sensitivities from above and put the relations for $G$ and $C$ leading to 
\begin{align*}
T &= \frac{Z_i^k\hat{G}Q_i^{-k}\hat{C}}{1+Z_i^k\hat{G}Q_i^{-k}\hat{C}} \overset{Z_i^k = X_i^kQ_i^{k}}{=} \frac{X_i^{k}\hat{G}\hat{C}}{1+X_i^{k}\hat{G}\hat{C}} = \frac{\hat{G}\hat{C}}{X_i^{-k}+\hat{G}\hat{C}},\\
S_\mathrm{y} &= \frac{1}{1+Z_i^k\hat{G}Q_i^{-k}\hat{C}} =  \frac{1}{1+X_i^{k}\hat{G}\hat{C}} \quad\text{and}\\
S_\mathrm{u} &=  \frac{Z_i^k\hat{G}}{1+Z_i^k\hat{G}Q_i^{-k}\hat{C}} = \frac{X_i^kQ_i^{k}\hat{G}}{1+X_i^{k}\hat{G}\hat{C}} =  \frac{Q_i^{k}\hat{G}}{X_i^{-k}+\hat{G}\hat{C}}.
\end{align*}
Knowing that all RHP-roots are part of $Z_i^k=X_i^kQ_i^{k}$, where $Q_i^{k}$ only has stable roots, it directly follows: The control loop is internally stable if all roots of the FO pseudo polynomial $X_i^{-k}+\hat{G}\hat{C}$ are located in $\mathbb{C}_\alpha$.
This also applies to the partial cancellation of a stable pair of conjugate complex poles, as $\mathcal{X}_{p,\alpha}$ only has stable roots. 
Furthermore, it holds for the implicit terms $\tilde{Q}_i^{k}$ and $ \tilde{X}_i^{k}$, utilizing the approximation proposed in the next section. 

To show the difference between partial cancellation of an unstable pole and the IO unstable pole-zero cancellation, consider the plant $G$, $k=-1$ and $i=1$, i.e. $G=Z_1^{-1}\hat{G}$. For the IO cancellation we use $C_\mathrm{IO} = Z_1\hat{C}_\mathrm{IO}$ leading to
\begin{align*}
S_{\mathrm{u,IO}} &=  \frac{Z_1^{-1}\hat{G}}{1+Z_1^{-1}\hat{G}Z_1\hat{C}} =  \frac{\hat{G}}{Z_1\left(1+\hat{G}\hat{C}\right)},
\end{align*}
where $S_{\mathrm{u,IO}}$ is unstable due to the RHP pole in the denominator, even if $T_\mathrm{IO}$ is stabilized by $C_\mathrm{IO}$. 
In contrast, considering the partial cancellation, we get
\begin{align*}
S_\mathrm{u} &= \frac{Q_{z,\nu}^{-1}\hat{G}}{X_{z,\alpha}+\hat{G}\hat{C}} =  \frac{\hat{G}}{Q_{z,\nu}\left(X_{z,\alpha}+\hat{G}\hat{C}\right)}.
\end{align*}
This implies that $S_\mathrm{u}$ is stable if and only if $T$ is stable which is in accordance with the relation above.

\section{Approximations of Fractional-Order Transfer-Functions}\label{sec:fo-oust}
Controllers with FO elements show to have advantages as a design tool. However, for implementation purposes in a real-time experimental setup with limited physical memory, the non-local operator~\eqref{eq:Caputo-definition} is to be approximated with IO terms to an arbitrary degree of precision (see~\citep{Li2016}), e.g. with
continued fraction expansion or Oustaloup filter (\cite{MonjeCVXF10}). The latter approximates the FO operator in a predefined frequency range ${\Omega = \{\omega \in \mathbb{R}\,|\,\omega_\mathrm{l}\leq \omega \leq \omega_\mathrm{h} \}}$ with an order $N$, leading to an IO approximation of order $N_{\mathrm{oust}}=2N+1$. It is given by (\cite{MonjeCVXF10})
\begin{equation}
	s^\alpha \approx H_\alpha(s)\! =\! \omega_\mathrm{h}^\alpha \!\! \prod_{k=-N}^N\!\frac{s+\omega_k^-}{s+\omega_k^+},\,\omega_k^\pm = \omega_\mathrm{l}\left(\frac{\omega_\mathrm{h}}{\omega_\mathrm{l}}\right) ^\frac{k+N+(1\pm\alpha)/2}{2N+1}
\end{equation}
with $\alpha\in (0,1)$, only consists of poles and zeros in $\mathbb{C}_-$ and therefore does not affect the open-loop design method.

An additional integrator is introduced to achieve the correct stationary gain, i.e. we use the approximation
\begin{equation}
s^{-\alpha}  = \frac{s^{1-\alpha}}{s}\approx\frac{H_{1-\alpha}(s)}{s},
\end{equation}
as applied e.g.~by \cite{Hosse13}.

In order to find an Oustaloup approximation for the implicit terms \citep{Oustaloup2000}, we use the substitution $\tilde{s} = 1-\frac{s}{z}$ resulting in
\begin{equation*}
\tilde{Q}_{z,\alpha}^{-1}(s) = \left(1-\frac{s}{z}\right)^{-\alpha}\!\!= \tilde{s}^{-\alpha}
\approx\frac{H_{1-\alpha}(\tilde{s})}{\tilde{s}} = \frac{H_{1-\alpha}\left(1-\frac{s}{z}\right)}{1-\frac{s}{z}}.
\end{equation*}

\begin{figure}[ht]
	\centering
	\includegraphics[width=\linewidth]{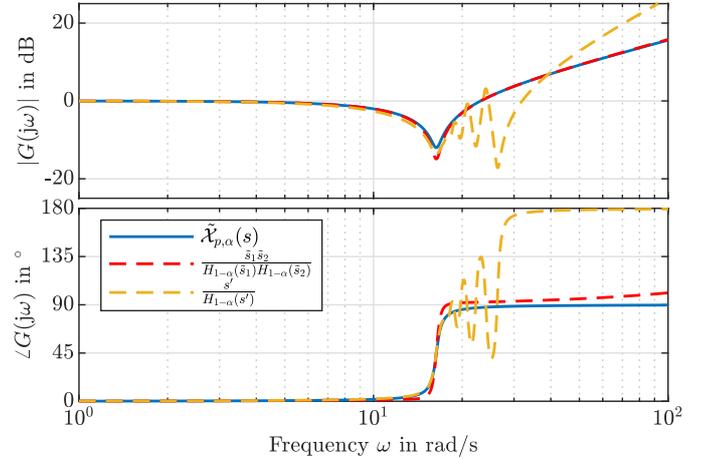}
	\caption{Oustaloup approximations of implicit term $\tilde{\mathcal{X}}_{p,\alpha}$ in the range $[\omega_l,\, \omega_h] = [0.5\,\mathrm{\frac{rad}{s}},\, 500\,\mathrm{\frac{rad}{s}}]$ with $N=5$.}
	\label{fig:implementation-Q-bode}
\end{figure}

For the second-order terms of the implicit representations $\tilde{\mathcal{Q}}_{z,\alpha}$ and $\tilde{\mathcal{X}}_{p,\alpha}$, introducing ${s^\prime = s^{2} - 2s\omega_0 \cos\left( \varphi\right) + \omega_0^{2}}$ would be natural. However, it turns out to be necessary to approximate each conjugate complex root individually
\begin{equation*}
	\tilde{\mathcal{X}}_{p,\alpha}(s) =  \underbrace{\left(1-\frac{s}{p}\right)^\alpha}_{ \tilde{s}_1^\alpha}~  \underbrace{\left(1-\frac{s}{\bar{p}}\right)^\alpha}_{\tilde{s}_2^\alpha}
	\approx \frac{\tilde{s}_1\tilde{s}_2}{H_{1-\alpha}(\tilde{s}_1)H_{1-\alpha}(\tilde{s}_2)}
\end{equation*}
with $p$ and $\bar{p}$ in~\eqref{eq:fo-def-p12} to achieve acceptable results, as shown  in Fig.~\ref{fig:implementation-Q-bode} for $p= -0.514+ \j 16.346 $ and $\alpha=0.5$.
Although $H_{1-\alpha}(\tilde{s}_1)$ has complex coefficients, the overall approximation only contains real-valued coefficients.

\section{Illustrative Example}
\label{sec:examples}
We consider a plant with a dominant non-minimum phase zero at $z=1$ given by
\vspace{-1.2ex}
\begin{equation}
G(s) =  \frac{s-1}{(1+\frac{s}{2})(1+\frac{s}{3})}
\end{equation}
and design four controllers to study the effect of the different compensation strategies:
\begin{align*}
C_1(s) &= k_1\frac{\tau s + 1}{\tau s}, &C_2(s) &= k_2\frac{\tau s + 1}{\tau s}D_1(s) \\
C_3(s) &= k_3\frac{\tau s + 1}{\tau s}Q_{z,2}(s), &C_4(s) &= k_4\frac{\tau s + 1}{\tau s} \tilde{Q}_{z,2}(s).
\end{align*}
All controllers consist of a classical PI controller with proportional gain $k_i, i=1,\ldots,4$, and time constant $\tau = 2$. Controller $C_1$ has no further component, whereas $C_2$ additionally pseudo compensates the non-minimum phase zero with the IO term $D_1$, compare~\eqref{eq:fo-def-D1}. Both $C_3$ and $C_4$ partially compensate $z$ with the explicit and implicit term $Q_{z,2}$ of~\eqref{eq:fo-exp-expansion} and $ \tilde{Q}_{z,2}$ of~\eqref{eq:fo-Q-imp}, respectively. For the comparison, all controllers are tuned to have an open-loop crossover frequency of $\omega_\mathrm{c}=0.54\,\mathrm{rad/s}$. As the (partial) compensation significantly affects the amplitude response, the proportional gains need individual adjustments resulting in $\begin{bmatrix}k_1,k_2,k_3,k_4\end{bmatrix} = \begin{bmatrix}0.68,0.772,1.091,0.7245\end{bmatrix}$. 

\begin{figure}[ht]
	\centering
	\includegraphics[width=1\linewidth]{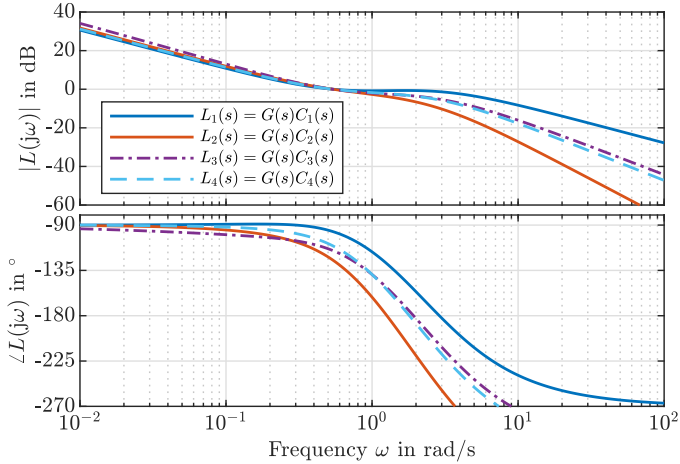}
	\caption{Bode plot of $L_i = GC_i$, $i=1,\ldots,4$.}
	\label{fig:examples-1-bode}
\end{figure}

A Bode plot of the open loops $L_i = GC_i$ is shown in Fig.~\ref{fig:examples-1-bode}. Obviously, the controllers cause a different amount of phase lag and slope at high frequencies (for the FO elements compare Table~\ref{tab:do-zero-asymptotics}). All open loops show a phase margin of $\Phi_{\mathrm{r},i}>55^\circ$. Also the gain margins are considered sufficient $A_{\mathrm{r},i}>3\,\mathrm{dB}$ for $i=2,3,4$, apart from $L_1$ with $A_{\mathrm{r},1}=1.26\,\mathrm{dB}$. The gain at low frequencies is almost identical, only $L_3$ is slightly higher. 

\begin{figure}[ht]
	\centering
	\includegraphics[width=1\linewidth]{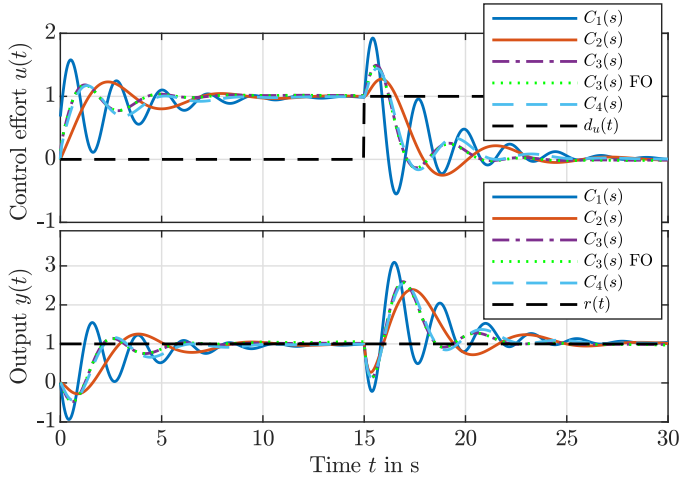}
	\caption{Closed-loop time-responses.}
	\label{fig:examples-1-step-r}
\end{figure}

To point out the differences between the controllers, the closed-loop step- and disturbance response are given in Fig.~\ref{fig:examples-1-step-r}. 
For these simulations, the FO elements $Q_{z,2}$ and $ \tilde{Q}_{z,2}$ are approximated with an Oustaloup filter in the range $[\omega_l,\, \omega_h] = [0.001\,\mathrm{\frac{rad}{s}},\, 1000\,\mathrm{\frac{rad}{s}}]$ with $N=5$. In addition to that, the results for $C_3$ are verified using an FO solver, illustrating the accuracy of the approximation.

Both step-responses with $C_1$ reveal the necessity to address the non-minimum phase zero for this tuning if such crossover frequency is to be achieved. Although an improvement of the tuning could further reduce the oscillations, the significant reduction of both the oscillations as well as the initial undershoot are caused by the additional controller elements. In comparison with the IO pseudo compensation, both controllers with FO elements show faster convergence and less overshoot for the reference step. In this case, the initial undershoot is slightly more visible, however still significantly less than without compensation. Compared to the implicit compensation with $C_4$, the controller $C_3$ with explicit partial cancellation leads to a faster convergence, due to the higher proportional gain to achieve the same crossover frequency~$\omega_\mathrm{c}$.

Also the time-response to a disturbance step in Fig.~\ref{fig:examples-1-step-r} shows similar behavior. The convergence with controllers $C_3$ and $C_4$ is still faster than with $C_2$, i.e. the IO pseudo compensation. However, the transient behavior is comparable with the controllers containing FO elements.

\section{Conclusions}
\label{sec:conclusions}
Applying the proposed methods to an example reveals the advantages of partially compensating RHP poles/zeros. 
It shows that utilizing the FO elements can reduce the undershoot for plants involving non-minimum phase zeros significantly. 
Furthermore, the FO elements introduce less additional phase lag compared to the IO pseudo compensation. If the non-minimum phase zero imposes phase limitations, the implicit term is preferred. 

Striving to counteract a low-damped pair of conjugate complex poles, partial compensation can be useful as well. The explicit partial cancellation completely erases the oscillatory behavior, whereas the implicit compensation only reduces the resonance peak. If the implicit representation is utilized, the Oustaloup approximation has to be carried out separately for each conjugate complex pseudo zero. \\
Future work will cover the relation between order $\alpha$ and time-domain overshooting as well as FO notch filter design.

\section*{Acknowledgment}
This research was supported by IS-DAAD program under RCN project no. 320067
(DAAD project-ID 57458791). 

\bibliography{literature-FO-control}

\end{document}